\documentclass[aps,twocolumn,showpacs,superscriptaddress,pra,10pt]{revtex4-2}

\usepackage{graphicx}
\usepackage{dcolumn}
\usepackage{bm}
\usepackage{hyperref}
\usepackage{tikz}
\usetikzlibrary{matrix}
\usepackage{xcolor}

\IfFileExists{newtxtext.sty}
    {\usepackage{newtxtext,newtxmath}}
    {\IfFileExists{stix.sty}
       {\usepackage{stix}}
       {\IfFileExists{mathptmx.sty}
       {\usepackage{mathptmx}}{} } }

\begin{document}

\preprint{APS/123-QED}

\title{
Adaptive quantum dynamics with the time-dependent variational Monte Carlo method
}

\author{Raffaele Salioni}
\email{raffaele.salioni@studenti.unimi.it}
\affiliation{Dipartimento di Fisica ``Aldo Pontremoli'', Universit\`a degli Studi di Milano, via Celoria 16, 20133 Milano, Italy}

\author{Rocco Martinazzo}
\email{rocco.martinazzo@unimi.it}
\affiliation{Dipartimento di Chimica, Universit\`a degli Studi di Milano, via Golgi 19, 20133 Milano, Italy}

\author{Davide Emilio Galli}
\email{davide.galli@unimi.it}
\affiliation{Dipartimento di Fisica ``Aldo Pontremoli'', Universit\`a degli Studi di Milano, via Celoria 16, 20133 Milano, Italy}

\author{Christian Apostoli}
\email{christian.apostoli@unimi.it}
\affiliation{Dipartimento di Fisica ``Aldo Pontremoli'', Universit\`a degli Studi di Milano, via Celoria 16, 20133 Milano, Italy}

\date{\today}

\begin{abstract}
We introduce an extension of the time-dependent variational Monte Carlo (tVMC) method that adaptively controls the expressivity of the variational quantum state during the simulation of the dynamics. This adaptive tVMC (atVMC) approach is specifically designed to enhance numerical stability when overparameterized variational ans\"atze lead to ill-conditioned equations of motion. Building on the concept of the local-in-time error (LITE)---a measure of the deviation between variational and exact evolution---we introduce a procedure to quantify each parameter's contribution to reducing the LITE, using only quantities already computed in standard tVMC simulations. These relevance estimates guide the selective evolution of only the most significant parameters at each time step, while maintaining a prescribed level of accuracy. We benchmark the algorithm on quantum quenches in the one-dimensional transverse-field Ising model using both spin-Jastrow and restricted Boltzmann machine wave functions, with an emphasis on overparameterized regimes. The adaptive scheme significantly improves numerical stability and reduces the need for strong regularization, enabling reliable simulations with highly expressive variational ans\"atze.
\end{abstract}

\maketitle

\section{\label{sec:intro}Introduction}

Understanding the dynamics of strongly correlated quantum many-body systems remains a fundamental and long-standing challenge in physics. In recent years, this problem has attracted renewed interest, driven by theoretical and experimental advances in the study of out-of-equilibrium phenomena~\cite{RevModPhys.83.863, PhysRevLett.105.015702, Heyl_2018, Zhang_time_crystal_2017, Choi_time_crystal_2017, Langen_2016, Abanin_2017, Bernien_Quantum_scars_2017}, as well as by the rapid development of quantum technologies~\cite{RevModPhys.86.153, Preskill2018quantumcomputingin, ezratty2024understandingquantumtechnologies2024}. In particular, accurately describing quantum dynamics is essential to understanding the propagation of entanglement and coherence, and ultimately, to achieving control over them~\cite{entanglement_review}.
Exact solutions to the time evolution of quantum many-body systems are limited to a few analytically tractable or numerically trivial cases. The primary obstacle is the exponential growth of the Hilbert space with the number of degrees of freedom, which makes a full description of the wave function computationally intractable. Consequently, a variety of numerical methods have been developed to circumvent this limitation, including tensor network techniques, quantum Monte Carlo, and variational approaches based on neural-network quantum states, all of which aim to efficiently parameterize the wave function in regimes where exact methods become intractable~\cite{SCHOLLWOCK201196_DMRG, orus2014practical, Orus2019, BeccaSorellaQuantumMC, Carleo_Solving_many-body_with_neural_networks, glasser2018neural}.

In this work, we focus on the time-dependent variational Monte Carlo (tVMC) method~\cite{Carleo_Becca_Schiró_Fabrizio_2012, Carleo_Becca_Sanchez_2014, Carleo_PhD_Thesis}, which approximates the system's wave function using a variational ansatz and, as described below, evolves it according to a variational principle. A key feature of tVMC is its flexibility in treating systems of arbitrary dimensionality, as its formulation does not rely on a specific geometry and connectivity.
Moreover, since the wave function is represented by a parameterized variational ansatz, the computational cost of the method depends on the number of variational parameters rather than on the dimension of the Hilbert space, allowing for favorable scaling even in large or high-dimensional systems.
This makes it complementary to other powerful methods, such as tensor networks~\cite{DMRG_PhysRevLett.69.2863,  SCHOLLWOCK201196_DMRG, orus2014practical, Orus2019, PAECKEL2019167998, 10.21468/SciPostPhysLectNotes.8, PhysRevLett.126.170603}, which are most effective in low-dimensional systems or when the entanglement is sufficiently constrained. However, the accuracy of a tVMC simulation crucially depends on the expressivity of the chosen variational ansatz.
Recently, highly expressive classes of variational wave functions based on machine learning techniques---commonly referred to as neural-network quantum states (NNQS)---have been introduced~\cite{Carleo_Solving_many-body_with_neural_networks}. While these ans\"atze can efficiently capture complex correlations~\cite{PhysRevB.106.205136, Ureña_Sojo_Bermejo-Vega_Manzano_2024, Dash_Gravina_Vicentini_Ferrero_Georges_2025, PhysRevX.7.021021} and have been successfully used to simulate the quantum many-body dynamics of several systems using tVMC~\cite{Carleo_Solving_many-body_with_neural_networks, PhysRevB.103.165103, doi:10.1126/sciadv.abl6850, PhysRevLett.131.046501, PRXQuantum.4.040302, PhysRevB.110.104411, PhysRevB.109.235120, PhysRevB.109.155128}, their application in this context often suffers from numerical instabilities~\cite{Hofmann_Role_of_stochastic_noise}. It is well known that this problem can be mitigated by the use of regularization techniques~\cite{Hofmann_Role_of_stochastic_noise, Schmitt_and_Heyl}, which, although effective in stabilizing the simulation, introduce biases that are difficult to quantify.
Addressing these issues is the main motivation of the present work.

To this end, we build on the concept of the local-in-time error (LITE), a quantity recently introduced in the context of variational dynamics~\cite{Martinazzo_LITE, martinazzo2021commentregularizingmctdhequations}, which can be evaluated within the tVMC framework and provides a local estimate of the accuracy of the variational evolution. The LITE provides an intrinsic estimate of the local accuracy of the time evolution at each simulation step. We employ this idea to develop an adaptive scheme---adaptive time-dependent variational Monte Carlo (atVMC)---which dynamically adjusts the number of variational parameters evolved during the tVMC simulation, based on their estimated relevance. 
Specifically, we derive an analytical criterion from the LITE to assess the importance of each parameter during the evolution.
This enables us to retain only those parameters contributing significantly to the dynamics, while keeping the squared LITE below a prescribed threshold, thus effectively and dynamically filtering out redundant expressivity in the variational manifold.
We benchmark the performance of our method on the one-dimensional transverse-field Ising (TFI) model, demonstrating that accurate simulations can be achieved while evolving only a subset of the variational parameters at each timestep. Moreover, we show that our adaptive strategy reduces the need for aggressive regularization, thereby improving overall stability and accuracy.

\section{\label{sec:methods}Methods}

In this section, we describe the theoretical and algorithmic framework developed in this work. We begin by reviewing the time-dependent variational Monte Carlo (tVMC) method, which provides the main foundation of our approach. We then introduce the local-in-time error (LITE), a quantity that measures the local accuracy of the variational time evolution and plays a central role in our adaptive scheme. Building upon these tools, we present the adaptive tVMC (atVMC) algorithm, which dynamically selects the subset of variational parameters to be updated at each time step of the tVMC simulation. The adaptive strategy aims to reduce the number of variational parameters that can evolve in time
while maintaining a controlled accuracy, as quantified by the LITE. 
Specific procedures for adding or removing parameters from the evolution are discussed in detail, together with a criterion to assess the significance of each parameter, which allows us to suppress directions in parameter space that do not contribute meaningfully to the physical dynamics and may lead to instabilities.

\subsection{\label{subsec:tVMC}The time-dependent variational Monte Carlo method}
The tVMC method employs a variational ansatz to represent the state of a quantum system, reducing the problem of describing its dynamics to determining the evolution of its time-dependent variational parameters~\cite{Carleo_PhD_Thesis, Carleo_Becca_Schiró_Fabrizio_2012, Carleo_Becca_Sanchez_2014}.
We denote by $\left | \psi (\boldsymbol{\alpha}_t) \right \rangle$ the variational ansatz for the quantum state and by $\psi (\boldsymbol{x}; \boldsymbol{\alpha}_t) = \langle \boldsymbol{x} | \psi(\boldsymbol{\alpha}_t) \rangle$ the variational wave function, where $\boldsymbol{x}$ is a generic configuration vector and $\boldsymbol{\alpha}_t$ is a vector of time-dependent variational parameters, $\alpha_k = \alpha_k(t) \in \mathbb{C}$.
The expectation value at time $t$ of any observable $\hat{O}$ can be obtained as
\begin{equation}
    \frac{\langle \psi(\boldsymbol{\alpha}_t) | \hat O | \psi(\boldsymbol{\alpha}_t) \rangle}{\langle \psi(\boldsymbol{\alpha}_t) | \psi(\boldsymbol{\alpha}_t) \rangle} = \sum_{\boldsymbol{x}} \frac{|\psi (\boldsymbol{x}; \boldsymbol{\alpha}_t)|^2}{\| \psi(\boldsymbol{\alpha}_t) \|^2} O_\text{loc}(\boldsymbol{x}; \boldsymbol{\alpha}_t) \text{,}
\end{equation}
where $O_\text{loc}(\boldsymbol{x}; \boldsymbol{\alpha}_t) = \sum_{\boldsymbol{x}'} \langle \boldsymbol{x} | \hat{O} | \boldsymbol{x}' \rangle \psi(\boldsymbol{x}';\boldsymbol{\alpha}_t) / \psi(\boldsymbol{x};\boldsymbol{\alpha}_t)$ is the \textit{local estimator} for the observable $\hat{O}$, and the expectation value can be calculated efficiently by Monte Carlo sampling the probability $p(\boldsymbol{x}; \boldsymbol{\alpha}_t) = |\psi (\boldsymbol{x}; \boldsymbol{\alpha}_t)|^2 / \| \psi(\boldsymbol{\alpha}_t) \|^2$, obtained from the squared modulus of the variational wave function.

The evolution of the parameters at time $t$ is determined by demanding that their time derivatives $\dot{\boldsymbol{\alpha}}_t$ minimize, to leading order in $dt$, the Fubini--Study distance~\cite{Carleo_Solving_many-body_with_neural_networks, Provost1980}
\begin{equation}\label{eq:FS_principle}
   \mathcal{D} \left ( \left | \psi (\boldsymbol{\alpha}_t + \dot{\boldsymbol{\alpha}}_t \, dt) \right \rangle , e^{-\frac{i}{\hbar}\hat{H} \, dt}\left | \psi (\boldsymbol{\alpha}_t) \right \rangle   \right )
\end{equation}
between the variationally-evolved state $\left | \psi (\boldsymbol{\alpha}_t + \dot{\boldsymbol{\alpha}}_t \, dt) \right \rangle$ and the local exact evolution $e^{-\frac{i}{\hbar}\hat{H} \, dt}\left | \psi (\boldsymbol{\alpha}_t) \right \rangle$. This approach is equivalent to applying McLachlan's variational principle~\cite{McLachlan, Yuan2019theoryofvariational} and yields
the following set of equations of motion for the variational parameters:
\begin{equation}\label{eq:linear_system}
    i \hbar \sum_{k'}S_{k,k'}\dot{\alpha}_{k'} = F_k \text{,}
\end{equation}
where $ S_{k,k'} = \langle \mathcal{O}_k^* \mathcal{O}_{k'} \rangle - \langle \mathcal{O}_k^* \rangle \langle \mathcal{O}_{k'} \rangle$ is the quantum geometric tensor and $F_k=\langle \mathcal{O}_k^* \mathcal{E} \rangle - \langle \mathcal{O}_k^*\rangle \langle \mathcal{E} \rangle$ is the force vector. These are defined in terms of the \textit{local operators} $\mathcal{O}_k ( \boldsymbol{x};\boldsymbol{\alpha}_t)  = \frac{\partial}{\partial \alpha_k} \ln \langle \boldsymbol{x} | \psi (\boldsymbol{\alpha}_t) \rangle$ and the \textit{local energy} $\mathcal{E} (\boldsymbol{x}; \boldsymbol{\alpha}_t) = \langle \boldsymbol{x} | \hat{H} | \psi (\boldsymbol{\alpha}_t) \rangle / \langle \boldsymbol{x} | \psi (\boldsymbol{\alpha}_t) \rangle$. 
Expectation values, denoted by $\langle \cdot \rangle$, are computed with respect to the probability $p(\boldsymbol{x}; \boldsymbol{\alpha}_t) = |\psi (\boldsymbol{x}; \boldsymbol{\alpha}_t)|^2 / \| \psi(\boldsymbol{\alpha}_t) \|^2$ introduced above.

The tVMC algorithm relies on Monte Carlo sampling of $p(\boldsymbol{x}; \boldsymbol{\alpha}_t)$ to compute the expectation values required for evaluating $S_{k,k'}$ and $F_k$. The linear system~\eqref{eq:linear_system} is then solved for the time derivatives $\dot{\boldsymbol{\alpha}}_t$, which in turn determine the values of the parameters at the following time step $\boldsymbol{\alpha}_{t+\delta t}$, via an integrator for ordinary differential equations.

It is well known, however, that the numerical solution of the linear system~\eqref{eq:linear_system} may lead to jump-like instabilities in the simulation. 
These instabilities are attributed to the amplification of the stochastic noise affecting the Monte Carlo averages. Such amplification occurs because the equations of motion~\eqref{eq:linear_system} are typically ill-conditioned, due to redundancy in the parameterization of the variational quantum state---a feature that is common in more expressive and capable ans\"atze that employ a large number of parameters~\cite{Hofmann_Role_of_stochastic_noise}. 
Typically, such instabilities are mitigated through the application of regularization techniques~\cite{Hofmann_Role_of_stochastic_noise,Schmitt_and_Heyl,quench_critical_point}. However, any form of regularization inevitably introduces some degree of bias into the dynamics, making it desirable to limit the strength of regularization in order to maintain stability with minimal distortion.

We remark that the problem of instabilities can be circumvented by employing a fundamentally different method, the projected time-dependent variational Monte Carlo (ptVMC)~\cite{Sinibaldi2023unbiasingtime, Gutierrez2022realtimeevolution, PhysRevA.108.022210, Nys_Pescia_Sinibaldi_Carleo_2024}, which avoids the ill-conditioned linear system altogether. However, this approach may entail a considerably higher computational cost~\cite{Gravina2025neuralprojected}. More recently, alternative strategies have been proposed that depart from conventional time-stepping schemes, relying instead on global-in-time variational formulations that optimize the entire quantum trajectory over a finite time interval, by minimizing a suitable cost function that enforces the Schrödinger dynamics~\cite{Sinibaldi2024time,VandeWalle_2025}. While these methods may offer improved stability and accuracy, they also require solving high-dimensional optimization problems, which may limit their applicability to long-time simulations or large systems.

In the present work, we focus on mitigating the instability problem while remaining within the standard time-local tVMC framework, aiming to preserve its computational efficiency and scalability.

\subsection{\label{subsec:LITE}The local-in-time error}

To develop our adaptive algorithm, we need a quantifiable measure of the error in a given variational approximation of the exact dynamics.
A natural choice is the Fubini--Study distance from Eq.~\eqref{eq:FS_principle}. We define the instantaneous error per unit time as
\begin{equation}\label{eq:FSerror}
    \varepsilon_{\text{FS}}(t) = \lim_{dt \rightarrow 0}\, \frac{1}{d t} \mathcal{D}(\left | \psi (\boldsymbol{\alpha}_t + \dot{\boldsymbol{\alpha}}_t\, dt) \right \rangle , e^{-\frac{i}{\hbar}\hat{H} \, dt}\left | \psi (\boldsymbol{\alpha}_t) \right \rangle) \text{.}
\end{equation}
This is a nonnegative quantity that characterizes the rate of departure between the locally exact solution and a state that evolves on the variational manifold. A somewhat lengthy but straightforward calculation (see the Supplemental material of Ref.~\cite{Schmitt_and_Heyl}) shows that $\varepsilon_{\text{FS}}^2$ can be expressed in terms of the energy variance $\operatorname{Var}(\hat{H})$, the quantum geometric tensor $S_{k,k'}$, the force vector $F_k$, and the time derivatives of the variational parameters $\dot{\boldsymbol{\alpha}}_k$. The result is
\begin{equation}\label{eq:FSerror_SH}
\begin{aligned}
 \hfill    \varepsilon_{\text{FS}}^2(t) &= \frac{1}{\hbar^2}\operatorname{Var}(\hat{H}) + \\
    &+ \sum_{k,k'} \dot \alpha_k^* S_{k,k'} \dot \alpha_{k'} + \frac{i}{\hbar} \sum_k \dot \alpha_k^* F_k - \frac{i}{\hbar} \sum_k F_k^* \dot \alpha_k \text{.}
\end{aligned}
\end{equation}
Importantly, Eq.~\eqref{eq:FSerror_SH} holds for arbitrary trajectories on the variational manifold and does not require that the parameter velocities $\dot{\alpha}_k$ satisfy the tVMC equations~\eqref{eq:linear_system}.
However, when the evolution of the variational state follows the tVMC equations of motion~\eqref{eq:linear_system}, $\varepsilon_{\text{FS}}^2$ attains its variational minimum and simplifies to a more compact form:
\begin{equation}\label{eq:LITE}
    \varepsilon^2(t) = \min_{\dot{\boldsymbol{\alpha}}} \varepsilon_{\text{FS}}^2(t) = \frac{1}{\hbar^2}\text{Var}(\hat{H}) - \sum_{k,k'} \dot \alpha_k^* S_{k,k'} \dot \alpha_{k'} \text{.} 
\end{equation}
Following the terminology of Refs.~\cite{Martinazzo_LITE, martinazzo2021commentregularizingmctdhequations}, we call $\varepsilon(t)$ the \textit{local-in-time error} (LITE).
Since $S$ is positive semi-definite (being a covariance matrix), and $\varepsilon^2$ is nonnegative by definition, from Eq.~\eqref{eq:LITE} it follows that $ 0 \leq \varepsilon^2(t) \leq \text{Var} (\hat{H})/\hbar^2$. Beyond serving as a local error diagnostic, the LITE also provides an \textit{a posteriori} upper bound to the global error~\cite{Lubich_2008}
\begin{equation}
\| \left | \psi (\boldsymbol{\alpha}_t) \right \rangle - \left | \psi_\text{ex} (t) \right \rangle \| \leq \int_0^t \varepsilon (t') \, dt' \text{,}
\end{equation}
where $\left | \psi_\text{ex} (t) \right \rangle$ is the exact quantum state at time $t$.
The analytic expressions in Eqs.~\eqref{eq:FSerror_SH} and~\eqref{eq:LITE} provide the theoretical basis for the development of our adaptive algorithm.

\subsection{\label{subsec:atVMC}Adaptive tVMC}

The algorithm we introduce in this work enables the on-the-fly adjustment of the number of variational parameters that are actively evolved during a tVMC simulation. A \textit{frozen}, or \textit{deactivated}, parameter remains constant during the time evolution, simplifying the equations of motion~\eqref{eq:linear_system} with the aim of improving their conditioning and, consequently, the numerical stability of the simulation.
Obviously, changing the ansatz expressivity affects the accuracy of the variational description of the evolution of the quantum state: freezing a parameter will in general increase the LITE.
Our goal is to use the smallest number of active parameters while ensuring that $\varepsilon^2(t)$, the LITE squared,  remains below a predefined error threshold $\lambda^2_{\text{LITE}}$. The decision to freeze or unfreeze parameters is based on their level of importance, defined in terms of their impact on the LITE. Specifically, we estimate the importance of a parameter by how much its presence reduces the error, thus allowing for a time-dependent ranking of the parameters based on their relevance. 
The rationale behind this approach is twofold. First, not evolving less important parameters reduces the risk of instabilities during the numerical treatment of their evolution. Second, as a beneficial by-product, reducing the number of parameters to be updated reduces the computational cost of solving Eq.(\ref{eq:linear_system}).
In the following, we detail how it is possible to accurately control, via the LITE estimation, the freezing or unfreezing of one or more parameters in the context of the tVMC method.

\subsubsection{Freezing one parameter}
At each time step, we compare the squared LITE, $\varepsilon^2(t)$, with the threshold, $\lambda^2_{\text{LITE}}$. The possibility of freezing a parameter is considered only if $\varepsilon^2(t) < \lambda^2_{\text{LITE}}$. In this case, we need to estimate the importance of each active parameter $\alpha_j$. For each $\alpha_j$ we perform a rearrangement of the parameter ordering so that $\alpha_j$ appears last. In the corresponding reordered basis, the $S$ matrix and the $F$ vector are expressed in block-matrix form as 
\begin{equation}
    S =
\begin{bmatrix}
    \tilde{S} & \boldsymbol{V}_{j} \\
    \boldsymbol{V}_{j}^{\dagger} & S_{j,j}
\end{bmatrix} \quad \text{and} \quad \boldsymbol{F} =
\begin{pmatrix}
    \tilde{\boldsymbol{F}} \\
    F_j
\end{pmatrix} \text{,}
\end{equation}
where the scalar entries $S_{j,j}$ and $F_j$ correspond, respectively, to the diagonal element of the $S$ matrix and the component of the $\boldsymbol{F}$ vector associated with $\alpha_j$, while the vector $\boldsymbol{V}_j$ contains the off-diagonal elements of $S$ that encode the correlations between $\alpha_j$ and the remaining parameters. The submatrix $\tilde{S}$ and the subvector $\tilde{\boldsymbol{F}}$ correspond, respectively, to the quantum geometric tensor and the force vector that would result if $\alpha_j$ were held fixed, or frozen. 
In the most recent time step, parameter $\alpha_j$ was treated as active, so the matrix $S$ has already been inverted as part of solving the tVMC equations~\eqref{eq:linear_system}. We now express its inverse $S^{-1}$ in the same reordered basis, using the same block-matrix notation introduced above:
\begin{equation}
    S^{-1} =
\begin{bmatrix}
    \tilde{K} & \boldsymbol{W}_{j} \\
    \boldsymbol{W}_{j}^{\dagger} & [S^{-1}]_{j,j}
\end{bmatrix}  \text{,}
\end{equation}
where we introduced the submatrix $\tilde{K}$, the diagonal element $[S^{-1}]_{j,j}$, and the off-diagonal vector $\boldsymbol{W}_{j}$.

We are now in a position to evaluate the effect on the variational dynamics if $\alpha_j$ were frozen.
The time derivatives of the remaining active parameters, denoted by $\dot{\boldsymbol{\alpha}}'$, are obtained by solving a reduced version of the linear system~\eqref{eq:linear_system}, in which $S$ and $F$ are replaced by $\tilde{S}$ and $\tilde{F}$, respectively. In matrix notation, this gives
\begin{equation}\label{eq:alpha_dot_primo_freezing}
    \dot{\boldsymbol{\alpha}}' = - \frac{i}{\hbar} \tilde{S}^{-1} \tilde{\boldsymbol{F}} \text{.}
\end{equation}
To compute $\tilde{S}^{-1}$ efficiently, we exploit the fact that the full inverse $S^{-1}$ has already been computed. Applying a standard inversion formula for block matrices~\cite{Noble_linearAlgebra}, we obtain
\begin{equation}
    \tilde{S}^{-1} = \tilde{K} - \frac{1}{[S^{-1}]_{j,j}} \boldsymbol{W}_{j}\boldsymbol{W}_{j}^{\dagger} \text{.}
\end{equation}

We can now compute the squared LITE that would result if $\alpha_j$ were frozen. We denote this quantity by $\varepsilon^2_j(t)$, and evaluate it using Eq.~\eqref{eq:LITE}, with $\tilde{S}$ and $\dot{\boldsymbol{\alpha}}'$ in place of $S$ and $\dot{\boldsymbol{\alpha}}$, respectively. 
Since freezing a parameter reduces the expressivity of the ansatz, $\varepsilon^2_j(t)$ is necessarily greater than the current squared error $\varepsilon^2(t)$ resulting from using all the active parameters. We define the corresponding increase in the squared LITE as
\begin{equation}\label{eq:importance_freezing}
    \Delta \varepsilon^2_j(t) = \varepsilon_j^2(t) - \varepsilon^2(t) \text{.}
\end{equation}
The quantity $\Delta \varepsilon^2_j(t)$ defines the \textit{importance} of the parameter $\alpha_j$.
After computing $\Delta \varepsilon^2_j(t)$ for all the active parameters, we rank them by their importance and identify the least important one---i.e. the parameter with the smallest $\Delta \varepsilon^2_j(t)$. This parameter is frozen only if its deactivation keeps the square LITE below $\lambda^2_{\text{LITE}}$. If this condition is satisfied, the parameter is deactivated, and the next time step of the tVMC algorithm proceeds with one fewer active parameter. Otherwise, the active parameter set remains unchanged, and the next time step continues with all current active parameters.

The calculation of $\Delta \varepsilon^2_j(t)$ described above is analytically exact, with the only source of error arising from stochastic fluctuations in the Monte Carlo estimates of $S$ and $F$. In what follows, we derive an alternative expression for $\Delta \varepsilon^2_j(t)$ that, while approximate, is considerably simpler and remains sufficiently accurate for use within our algorithm.
The core approximation we introduce is to assume that deactivating the parameter $\alpha_j$ does not affect the time derivatives of the remaining parameters. In other words, we assume that the dynamics is not altered by the removal of $\alpha_j$, which is justified in our context, as we aim to freeze only those parameters that are deemed unimportant. Since this approximate dynamics no longer strictly follows the tVMC equations~\eqref{eq:linear_system}, we evaluate the resulting error per unit time $\varepsilon^2_{\text{FS},j}(t)$ using Eq.~\eqref{eq:FSerror_SH}, which does not depend on this assumption. The approximate importance of the parameter $\alpha_j$ is then given by the change in this quantity,
\begin{equation}\label{eq:Delta_FS_error}
\begin{split}
    \Delta \varepsilon_{\text{FS}, j}^2(t) &= \varepsilon_{\text{FS}, j}^2(t) - \varepsilon^2(t) \\
    &= -\dot{\alpha}^*_j \sum_{k \neq j}S_{j,k}\dot{\alpha}_{k} - \left( \sum_{k \neq j}\dot{\alpha}^*_{k}S_{k,j} \right)\dot{\alpha}_j \\
    &\qquad\qquad\quad + \dot{\alpha}_j^*S_{j,j}\dot{\alpha}_j - \frac{i}{\hbar}\dot{\alpha}^*_j F_j + \frac{i}{\hbar}F^*_j\dot{\alpha}_j \text{,}
\end{split}
\end{equation}
where $\dot\alpha_j$ is the current time derivative of the parameter under consideration, just before it is frozen. Under our assumption that the parameter derivatives other than $\dot\alpha_j$ are unchanged, we can substitute Eq.~\eqref{eq:linear_system} into Eq.~\eqref{eq:Delta_FS_error}, obtaining a very simple approximate expression for the variation of the LITE,
\begin{equation}
    \Delta \varepsilon^2_j(t) \simeq S_{j,j} \left | \dot{\alpha}_j \right |^2 \text{.}
\end{equation}
We note that this formula provides an overestimation of the true error increase, since if $\alpha_j$ were frozen, the derivatives of the remaining parameters would satisfy the reduced set of tVMC equations, namely Eq.~\eqref{eq:alpha_dot_primo_freezing}, and the quantity $\varepsilon^2_{\text{FS},j}$ would attain the variational minimum.

\subsubsection{Unfreezing one parameter}

If, at a given time $t$, the squared LITE $\varepsilon^2(t)$ exceeds the threshold $\lambda^2_{\text{LITE}}$, then we unfreeze one of the currently deactivated parameters to reduce the error below this limit. To select the most relevant parameter for reactivation, we evaluate the importance of each deactivated parameter $\alpha_l$ by comparing the current LITE with the LITE that would result from reactivating $\alpha_l$.

At the most recent time step, the tVMC equations~\eqref{eq:linear_system} have already been solved for the currently active parameters, yielding their time derivatives $\dot{\boldsymbol{\alpha}} = -(i/\hbar)S^{-1}\boldsymbol{F}$. To assess the impact of reactivating $\alpha_l$, we now introduce an extended quantum geometric tensor $\bar{S}$ and force vector $\bar{F}$, which include the entries associated with the parameter under investigation $\alpha_l$:
\begin{equation}
    \bar{S} =
    \begin{bmatrix}
        S & \bar{\boldsymbol{V}}_{l} \\
        \bar{\boldsymbol{V}}_{l}^{\dagger} & \bar{S}_{l,l}
    \end{bmatrix} \quad \text{and} \quad \bar{\boldsymbol{F}} =
    \begin{pmatrix}
        \boldsymbol{F} \\
        \bar{F}_l
    \end{pmatrix} \text{,}
\end{equation}
where the matrix $S$ and the vector $\boldsymbol{F}$ are, respectively, the quantum geometric tensor and the force vector corresponding to the currently active parameters. The scalar entries $\bar{S}_{l,l}$ and $\bar{F}_l$ denote the diagonal element of the $\bar{S}$ matrix and the component of the $\bar{\boldsymbol{F}}$ vector associated with $\alpha_l$. The vector $\bar{\boldsymbol{V}}_l$ contains the off-diagonal elements of $\bar{S}$ that encode the correlations between $\alpha_l$ and the already active parameters.
The inclusion of the additional entries modifies the tVMC equations and, consequently, the dynamics of all currently active parameters. We denote the updated parameter derivatives by
\begin{equation}
    \dot{\boldsymbol{A}} =
    \begin{pmatrix} 
        \dot{\boldsymbol{\alpha}}' \\
        \dot{\alpha}_l
    \end{pmatrix}
    = -\frac{i}{\hbar}\bar{S}^{-1} \bar{\boldsymbol{F}} \text{,}
\end{equation}
where $\dot{\boldsymbol{\alpha}}'$ represents the updated time derivatives of the previously active parameters after the hypothetical reactivation of $\alpha_l$. The extended tVMC system can be written in matrix form as $\bar{S}\dot{\boldsymbol{A}} = - (i/\hbar)\bar{\boldsymbol{F}}$. Using the block decomposition of $\bar{S}$, $\bar{\boldsymbol{F}}$, and $\dot{\boldsymbol{A}}$, this system is equivalently expressed as
\begin{equation}\label{system_equations_alpha_k}
    \begin{cases}
        S\dot{\boldsymbol{\alpha}}' + \bar{\boldsymbol{V}}_{l} \dot{\alpha}_l = -\frac{i}{\hbar}\boldsymbol{F} \\
        \bar{\boldsymbol{V}}_{l}^{\dagger} \dot{\boldsymbol{\alpha}}' + \bar{S}_{l,l}\dot{\alpha}_l = -\frac{i}{\hbar}\bar{F}_l
    \end{cases}
\end{equation}
Solving the first equation for $\dot{\boldsymbol{\alpha}}'$ yields
\begin{equation}\label{alpha_dot_primo}
    \dot{\boldsymbol{\alpha}}' = \dot{\boldsymbol{\alpha}} - S^{-1} \bar{\boldsymbol{V}}_{l} \dot{\alpha}_l \text{.}
\end{equation}
Substituting this expression into the second equation and solving for $\dot{\alpha}_l$ gives
\begin{equation}\label{alpha_dot_k_unfreezing}
    \dot{\alpha}_l = \frac{1}{\left ( \bar{S}_{l,l} - \bar{\boldsymbol{V}}_{l}^{\dagger} S^{-1} \bar{\boldsymbol{V}}_{l} \right )} \left( -\frac{i}{\hbar} \bar{F}_l - \bar{\boldsymbol{V}}_{l}^{\dagger} \dot{\boldsymbol{\alpha}} \right ) \text{.}
\end{equation}
To compute the LITE that would result from reactivating $\alpha_l$, we evaluate Eq.~\eqref{eq:LITE}, substituting the updated quantities $\bar{S}$ and $\dot{\boldsymbol{A}}$ for $S$ and $\dot{\boldsymbol{\alpha}}$. Using Eqs.~\eqref{alpha_dot_primo} and~\eqref{alpha_dot_k_unfreezing}, this quantity can be fully expressed in terms of $\dot{\boldsymbol{\alpha}}$, $S^{-1}$, $\bar{F}_l$, $\bar{S}_{l,l}$, and $\bar{\boldsymbol{V}}_{l}$. The resulting expression for the change in the squared LITE is
\begin{equation}\label{eq:importance_unfreezing}
    \Delta \varepsilon^2_l = \frac{1}{\bar{S}_{l,l} - \bar{\boldsymbol{V}}_{l}^{\dagger}S^{-1}\bar{\boldsymbol{V}}_{l}} \left | -\frac{i}{\hbar}\bar{F}_l - \bar{\boldsymbol{V}}_{l}^{\dagger}\dot{\boldsymbol{\alpha}}\right |^2 \text{.}
\end{equation}
Thus, after computing Eq.~\eqref{eq:importance_unfreezing} for all currently deactivated parameters, we reactivate the one with the largest $\Delta \varepsilon^2_l$.

Equation~\eqref{eq:importance_unfreezing} shows that, in addition to the quantities $S$ and $\boldsymbol{F}$---which are already required to evolve the active parameters---we must also compute, for each deactivated parameter $\alpha_l$, the corresponding force component $\bar{F}_l$, the diagonal element $\bar{S}_{l,l}$, and the off-diagonal vector $\bar{\boldsymbol{V}}_{l}$, which quantifies the coupling between the dynamics of $\alpha_l$ and those of the already active parameters.

\subsubsection{Collective parameter update}\label{subsubsec:collective_update}
If the squared LITE $\varepsilon^2(t)$ is significantly smaller than the threshold $\lambda^2_{\text{LITE}}$, it may be possible to freeze multiple parameters while still maintaining the error below the prescribed limit. Conversely, if $\varepsilon^2(t)$ is significantly larger than $\lambda^2_{\text{LITE}}$, unfreezing only one parameter may not suffice to reduce the error below the threshold. For such cases, we employ schemes that allow multiple parameters to be frozen or unfrozen simultaneously. We refer to the simultaneous activation or deactivation of multiple parameters in a single time step as a \textit{collective parameter update}.

\textit{Freezing $M$ parameters} --- When $\varepsilon^2(t)$ is well below the threshold $\lambda^2_{\text{LITE}}$, we consider deactivating several parameters at the same time. To identify suitable candidates, we compute the importance $\Delta \varepsilon_j^2(t)$ of each currently active parameter using Eq.~\eqref{eq:importance_freezing}, and sort the parameters in ascending order of importance, i.e., from least to most relevant.

An approximate estimate of the total error increase associated with freezing a group of $M$ parameters is obtained by summing the individual contributions:
\begin{equation}\label{eq:collective_update_freezing}
    \Delta \varepsilon^2_{\text{collective}} \simeq \sum_{j \in \mathcal{I}_M} \Delta \varepsilon_j^2(t)\text{,}
\end{equation}
where $\mathcal{I}_M$ denotes the set of indices corresponding to the $M$ least important parameters. This approximation neglects inter-parameter correlations and assumes the impact of freezing each parameter is independent. While this simplification may overestimate the total error, it provides an efficient and practical means of evaluating whether a group of parameters can be deactivated simultaneously. We then determine the largest number $M$ for which the estimated increase in error keeps the LITE below the threshold, and freeze all $M$ selected parameters simultaneously in a single collective update.

For situations that demand a more precise selection of parameters to freeze, one can employ the following alternative---and computationally more demanding---estimation method. A binary search is performed to identify the largest number $M$ of parameters that can be frozen without exceeding the error threshold. At each iteration $k$, the $M_k$ least relevant parameters in our ordered list are considered for reactivation. Then, the tVMC equations~\eqref{eq:linear_system} are solved, restricting the quantum geometric tensor $S$ and force vector $F$ to the subset of parameters that remain active. This procedure allows estimation of the resulting $\varepsilon^2(t)$ when those $M_k$ parameters are frozen. If this estimate exceeds the threshold $\lambda^2_{\text{LITE}}$, $M_k$ is decreased; otherwise, it is increased. The binary search proceeds until the maximal number of parameters that can be safely deactivated---while maintaining the error below the threshold---is found.

This approach involves multiple inversions of submatrices of $S$ and is therefore computationally more demanding. Nonetheless, the approximate estimate given by Eq.~\eqref{eq:collective_update_freezing} can be used as an efficient preliminary filter to assess the feasibility of freezing multiple parameters simultaneously. The more precise method is then employed only when this initial estimate indicates a potential benefit---a situation that occurs infrequently. We find this approach particularly useful at the very first time step of the simulation, where the algorithm begins with all parameters active and uses an initial collective freezing step to select the appropriate subset effectively.

It is worth noting that the binary search scales logarithmically (base $2$) with the number of currently active parameters, rendering it relatively efficient. However, if the computational cost remains prohibitive in practice, one can always fall back on the approximate method described earlier, which offers a much faster, though less precise, alternative.

\textit{Unfreezing $M$ parameters} --- When $\varepsilon^2(t)$ is well above the threshold $\lambda^2_{\text{LITE}}$, we consider activating multiple parameters simultaneously. We estimate the importance of each frozen parameter using Eq.~\eqref{eq:importance_unfreezing}, and sort them in descending order of importance, from the most to the least relevant.

We then iteratively subtract the estimated contributions $\Delta \varepsilon^2_l$ from the current value of $\varepsilon^2(t)$, following the sorted list, until the estimated error falls below the threshold. The corresponding top $M$ parameters are then selected for activation. This estimate of the error reduction is approximate, as it neglects correlations between the deactivated parameters.

In principle, one could account for such correlations by computing the relevant entries of the quantum geometric tensor involving pairs of deactivated parameters. This would allow solving the tVMC equations in an extended basis that includes the parameters under consideration for reactivation, providing a more accurate estimate of the resulting LITE. However, this would require inverting the $S$ matrix in a basis that includes currently deactivated and potentially irrelevant parameters, which could reintroduce the numerical instabilities we aim to avoid. For this reason, we favor the simpler approach for its greater stability and computational efficiency, and find that the approximation yields satisfactory results in practice.

\subsubsection{Preventing overparameterization}\label{subsubsec:significance}
In certain situations, the expressivity of the ansatz may be insufficient to keep the squared LITE below the threshold $\lambda^2_{\text{LITE}}$, even when all the variational parameters are active. However, not all of these parameters may contribute meaningfully to the dynamics; some may correspond to redundant or physically irrelevant degrees of freedom. The inclusion of such parameters in the evolution can induce numerical instabilities caused by overparameterization~\cite{Hofmann_Role_of_stochastic_noise}.

To identify and deactivate irrelevant parameters, one can evaluate the significance of each parameter $\alpha_k$ by its individual contribution to the global error, quantified by $\Delta \varepsilon^2_k$. A parameter is considered non-relevant if its contribution satisfies $\Delta \varepsilon^2_k < \eta_{\text{sig}}^2 \cdot \varepsilon^2(t)$, where $\varepsilon^2(t)$ is the total LITE squared at time $t$, and $\eta_{\text{sig}}^2 < 1$ is a user-defined \textit{significance threshold}. Parameters identified as non-relevant remain frozen even when the total squared LITE $\varepsilon^2(t)$ exceeds the threshold $\lambda^2_{\text{LITE}}$.

Additionally, the relevance of each parameter is reassessed at every time step. If one or more previously active parameters become non-relevant, the least significant among them is frozen. This dynamic filtering ensures that the set of active parameters remains focused on those that meaningfully contribute to reducing the variational error, thereby improving stability and efficiency in overparameterized regimes.

\section{\label{sec:results}Results}

To evaluate the effectiveness of our atVMC algorithm, we benchmark it on the one-dimensional transverse-field Ising model (TFI) with periodic boundary conditions. This model, which is exactly solvable~\cite{PFEUTY1970}, is defined by the Hamiltonian
\begin{equation}\label{eq:H_Ising}
  \hat{H} = - J \sum_i \hat{\sigma}_{i}^z \hat{\sigma}_{i+1}^z - h \sum_i \hat{\sigma}_i^x \text{,}
\end{equation}
where the index $i$ labels the spin sites, and $\hat{\sigma}_i^x$ and $\hat{\sigma}_i^z$ are Pauli operators.
The model is characterized by the dimensionless parameter $g = |h/J|$ and exhibits a quantum critical point at $g = g_c \equiv 1$, which marks a quantum phase transition: for $g > g_c$ the ground state is paramagnetic, while for $g < g_c$ it is ferromagnetic.
To explore nontrivial dynamics, we perform tVMC simulations of \emph{quantum quenches}. In a quantum quench, the system is initially prepared in the ground state corresponding to some value $g = g_1$ (the ground state in our simulations is variationally optimized via stochastic reconfiguration~\cite{Becca_Sorella_Optimization_Chapter, Sorella_stoc_recnfig}), and then evolved with a different parameter $g = g_2$. We refer to this protocol as the quench $g = (g_1 \to g_2)$. Simulating quenches where $g_2$ is near the critical point $g_c$ is particularly challenging, due to the expected growth of quantum correlations~\cite{quench_critical_point}.

We test atVMC on two commonly employed variational wave functions. The first one is the spin-Jastrow ansatz~\cite{Spin_Jastrow_state} with complex-valued parameters, defined as
\begin{equation}
\psi(\boldsymbol{\sigma}; \boldsymbol{\alpha}_t) = \exp \left( \sum_{i < j} \alpha_{i,j} \, \sigma_i \sigma_j \right) \text{,}
\end{equation}
where $\boldsymbol{\sigma} = (\sigma_1,\dots,\sigma_N)$ denotes a spin configuration of the $N$-site chain in the computational basis, and each parameter $\alpha_{i,j}$ encodes the correlation between spins at sites $i$ and $j$. Exploiting translational invariance and spatial reflection symmetry in our one-dimensional system, each $\alpha_{i,j}$ depends only on the distance between sites $i$ and $j$ (with periodic boundary conditions), reducing the number of independent parameters to $\lfloor N/2 \rfloor$.

The second variational ansatz we consider is the restricted Boltzmann machine (RBM) wave function as introduced in Ref.~\cite{Carleo_Solving_many-body_with_neural_networks}, which is defined as
\begin{equation}
\psi(\boldsymbol{\sigma}; \boldsymbol{\alpha}_t) = e^{\sum_i^{N}a_i \sigma_i}\prod_{j=1}^{Nd} \cosh \left(b_j + \sum_{i=1}^N W_{j,i} \sigma_i\right) \text{,}
\end{equation}
where the variational parameters $\boldsymbol{\alpha}_t$ consist of the visible biases $a_i$, hidden biases $b_j$, and the weight matrix $W_{j,i}$. The number of hidden units is $Nd$, where $d$ is the hidden-variable density. We adopt the translation-invariant version of the RBM, in which symmetry constraints require the density $d$ to be integer-valued and reduce the number of independent parameters to $Nd + d + 1$.

All simulation results presented in this work are freely available online~\cite{salioni_2025_15736912}.

\subsection{Spin-Jastrow wave function, quench $g= (4 \to 2)$}
To illustrate the behavior of the atVMC algorithm, we begin by considering the quench $g = (4 \to 2)$ in a TFI chain of $32$ spins. The system starts in a paramagnetic initial state and evolves under a different but still paramagnetic Hamiltonian. We employ the spin-Jastrow variational wave function, which in this case has $16$ independent parameters, and impose a LITE threshold of $\lambda^2_\text{LITE} = 10^{-2} \cdot \operatorname{Var}(\hat{H})/\hbar^2$. The tVMC equations of motion are solved using the pseudoinverse regularization with a tolerance of $10^{-7}$~\cite{Carleo_Solving_many-body_with_neural_networks,quench_critical_point,Schmitt_and_Heyl, Hofmann_Role_of_stochastic_noise}. Here, we employ the atVMC algorithm in its simplest form, allowing at most one parameter to be frozen or unfrozen per time step.

\begin{figure}
    \centering
    \includegraphics[width=\linewidth]{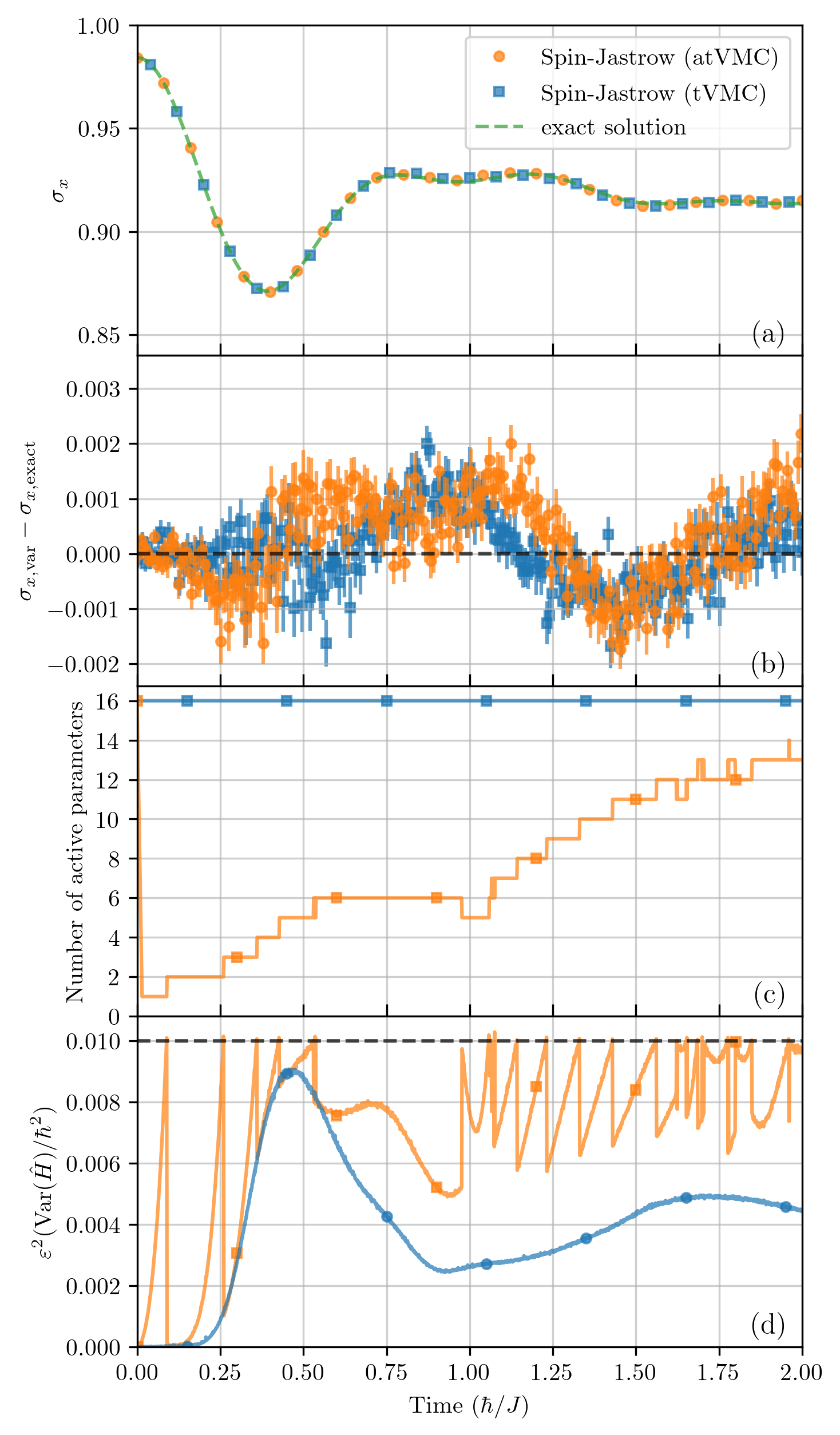}
    \caption{Quench $g = (4 \to 2)$ simulated using atVMC (circular markers, orange color) and tVMC (square markers, blue color) with a spin-Jastrow ansatz on a TFI chain of $32$ spins. (a) Transverse magnetization per spin: variational results compared with the exact solution (dashed line). Error bars are omitted as they are comparable in magnitude to the observable's oscillations.
    (b) Difference between the variational and the exact transverse magnetization per spin.
    (b) Number of active variational parameters. (c) Squared LITE during the evolution, compared to the target threshold (horizontal dashed line). }
    \label{fig:jas_4to2}
\end{figure}

Fig.~\ref{fig:jas_4to2}(a) shows the time evolution of the magnetization along the $x$-axis, defined as $\sigma_x = \left\langle \sum_i \hat{\sigma}^x_i \right\rangle / N$.
We compare the result obtained via atVMC with the exact solution and with the result obtained via standard tVMC, where all the parameters of the variational ansatz are active. The variational results are in very good agreement with the exact transverse magnetization and confirm that atVMC does not compromise accuracy when the ansatz is already well-conditioned.
In order to better appreciate the level of agreement, Fig.~\ref{fig:jas_4to2}(b) shows the difference of the transverse magnetization per spin computed with the Spin-Jastrow variational wave function and the exact value.
Fig.~\ref{fig:jas_4to2}(c) shows the number of active parameters as a function of time. All parameters are initially active, but during the first steps nearly all are quickly frozen: quantum correlations generated by the quench spread locally at first, and only the parameter encoding nearest-neighbor correlations needs to remain active. As the correlations propagate further through the system, the parameter describing next-nearest-neighbor correlations is the first to be unfrozen, followed by those corresponding to distance three, then four, and so on. At longer times, the pattern becomes more complex, with some medium-range parameters intermittently frozen.
Fig.~\ref{fig:jas_4to2}(d) plots $\varepsilon^2 (t)$, i.e., the squared LITE over time. Comparing it with panel (c), we observe that whenever $\varepsilon^2$ exceeds the threshold $\lambda^2_{\text{LITE}}$, a parameter is unfrozen, resulting in a sudden drop in the error which, interestingly, aligns with the error of the tVMC simulation. Conversely, when $\varepsilon^2$ is below $\lambda^2_{\text{LITE}}$, the algorithm freezes a parameter if doing so keeps the LITE within bounds. In these cases, $\varepsilon^2$ exhibits a sharp rise.

\subsection{RBM wave function, quench $g=(4 \to 1.5)$}
To illustrate the importance of the \textit{collective parameter updates} introduced in Sec.~\ref{subsubsec:collective_update}, we now consider the quench $g=(4 \to 1.5)$ in the $32$-spin TFI chain. This quench drives the system closer to the quantum critical point and generates more intricate quantum correlations, so we use a more expressive variational ansatz: a restricted Boltzmann machine (RBM) with density $d=3$, corresponding to $100$ variational parameters in our setup. We set the LITE threshold to $\lambda^2_\text{LITE} = 10^{-3} \cdot \operatorname{Var}(\hat{H})/\hbar^2$ and employ pseudoinverse regularization with a tolerance of $10^{-7}$.
In addition, we adopt the adaptive time-step scheme for the Heun integrator introduced in Ref.~\cite{Schmitt_and_Heyl}, which is compatible with atVMC upon a minor modification: the number of active parameters is updated only once every two steps.

\begin{figure}
    \centering
    \includegraphics[width=\linewidth]{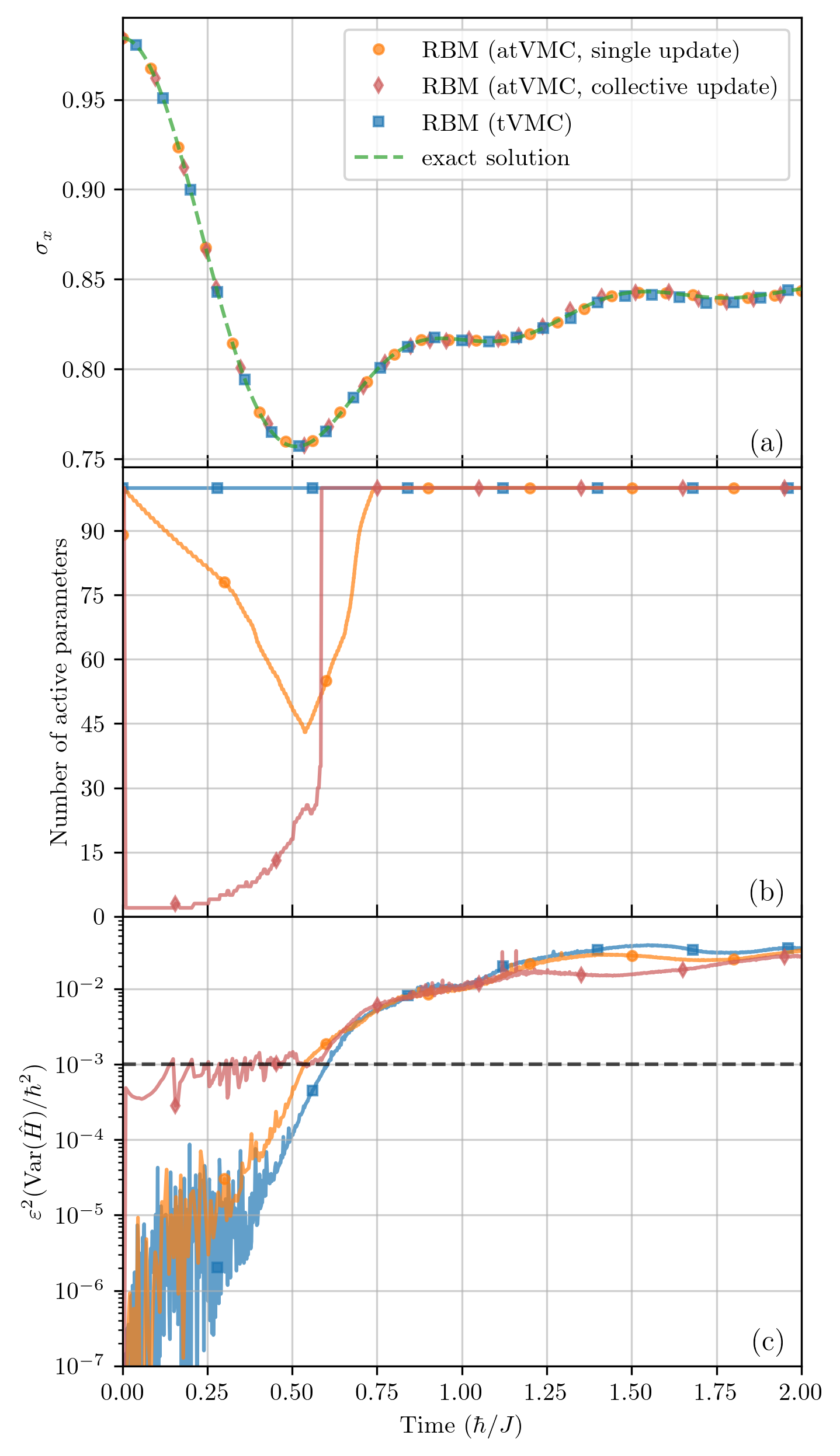}
    \caption{Quench $g=(4 \to 1.5)$ simulated using atVMC and tVMC with a RBM ansatz with density $d=3$ on a TFI chain of $32$ spins. atVMC simulations are performed with an adaptive time-step scheme, both with (diamond markers, red color) and without (circular markers, orange color) collective parameter updates. (a) Transverse magnetization per spin: variational results compared to the exact solution (dashed line). Error bars are omitted as they are smaller than the marker size. (b) Number of active variational parameters. (c) Squared LITE during the evolution, compared to the target threshold (horizontal dashed line).}
    \label{fig:RBM_4to1.5}
\end{figure}

Fig.~\ref{fig:RBM_4to1.5} compares atVMC simulations---with and without the collective parameter update---and a standard tVMC simulation, in which all the variational parameters of the ansatz are evolved. Panel (a) plots the evolution of the transverse magnetization $\sigma_x$, showing that the dynamics of the atVMC simulations is in quite good agreement with the exact result and the tVMC simulation.
Panels (b) and (c) show the number of active parameters and the squared LITE as functions of time, respectively. As in the previous quench, the algorithm initially wants only a small subset of parameters  to be active. At later times, however, all parameters are eventually unfrozen, since the full expressivity of the RBM with $d = 3$ is insufficient to keep the error within the imposed bound.

This scenario was deliberately chosen to illustrate the importance of the collective parameter update. Without it, the algorithm does not behave as intended: the rate at which parameters are frozen or unfrozen becomes dependent on the time-step size. As shown in Panel (b) of Fig.~\ref{fig:RBM_4to1.5}, the simulation without the collective update struggles to adjust the number of active parameters appropriately, and Panel (c) reveals that the LITE fails to stabilize at the target threshold.

By contrast, the simulation with the collective update immediately reaches the minimal number of parameters required to satisfy the LITE condition. Furthermore, once the error can no longer be reduced below the threshold, all remaining parameters are promptly unfrozen. This results in more consistent and reliable control over the LITE.

\subsection{RBM wave function, quench $g=(0.5 \to 1)$ }
We now consider the quench $g=(0.5 \to 1)$ in the 32-spin TFI chain. The initial ground state, prepared in the ferromagnetic phase, evolves under a Hamiltonian at the quantum critical point, so that simulating the quantum correlations generated by the quench is especially challenging~\cite{quench_critical_point}. In this case, we employ a particularly expressive ansatz, a RBM with density $d=15$, corresponding to $496$ variational parameters in our setup. We set the LITE threshold to $\lambda^2_\text{LITE} = 10^{-3} \cdot \operatorname{Var}(\hat{H})/\hbar^2$, and employ pseudoinverse regularization with a tolerance of $10^{-7}$.

\begin{figure}
    \centering
    \includegraphics[width=\linewidth]{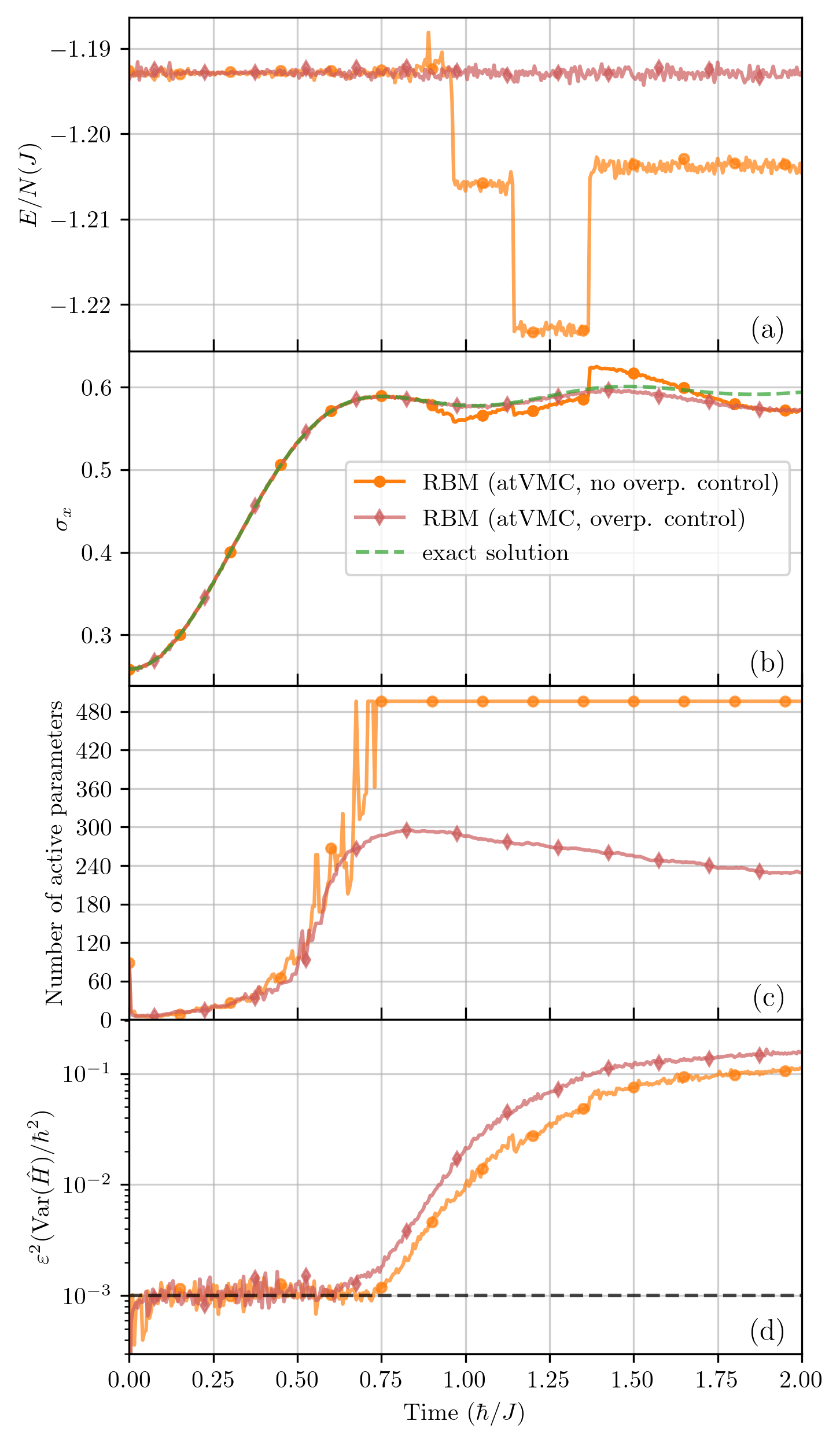}
    \caption{Quench $g=(0.5 \to 1)$ simulated using atVMC with a RBM ansatz with density $d=15$ on a TFI chain of $32$ spins, with collective parameter updates, with (diamond markers, red color) and without (circular markers, orange color) overparameterization control. (a) Energy per spin. (b) Transverse magnetization per spin: atVMC results compared to the exact solution (dashed line).  Error bars are omitted as they are comparable in magnitude to the observable's oscillations. (c) Number of active variational parameters selected by the adaptive algorithm. (d) Squared LITE during the evolution, compared to the target threshold (horizontal dashed line).}
    \label{fig:RBM_0.5to1}
\end{figure}

The curve with circular markers in Fig.~\ref{fig:RBM_0.5to1}(b) shows the evolution of the transverse magnetization $\sigma_x$ for an atVMC simulation that employs collective parameter updates but no mechanism to prevent overparameterization. In this case, the adaptive adjustment of the number of active parameters is insufficient to avoid instabilities. Such instabilities are clearly visible in Fig.~\ref{fig:RBM_0.5to1}(a) where the evolution of the energy per spin is shown: in any stable tVMC or atVMC simulation, the energy is conserved throughout the evolution; thus, the presence of discontinuities, as observed here,  provides a clear signature of numerical instabilities.
This is due to the difficulty of the quench: after a certain time, even the full expressivity of the ansatz cannot keep the LITE below the threshold. As shown in Fig.~\ref{fig:RBM_0.5to1}(c), the algorithm responds by activating all available parameters in an attempt to reduce the error. However, this also reintroduces many irrelevant parameters, which contributes to the onset of instabilities.

This issue is addressed by the overparameterization prevention scheme described in Sec.~\ref{subsubsec:significance}. The curve with triangular markers in Fig.~\ref{fig:RBM_0.5to1}(b) shows the evolution of $\sigma_x$ for an atVMC simulation that incorporates this control mechanism with significance threshold $\eta_{\text{sig}}^2 = 5 \cdot 10^{-3}$, which successfully prevents instabilities, as further confirmed by the good energy conservation shown in Fig.~\ref{fig:RBM_0.5to1}(a). The transverse magnetization remains compatible with the exact dynamics up to a certain point in the simulation, beyond which deviations arise due to the challenging nature of this quench. As shown in Fig.~\ref{fig:RBM_0.5to1}(c), even when the LITE exceeds the threshold, not all parameters are activated---an outcome of the overparameterization control scheme. Fig.~\ref{fig:RBM_0.5to1}(d) compares the squared LITE for the two simulations---with and without this control mechanism---and shows that, while the control scheme yields a slightly higher LITE, it remains of the same order of magnitude, while achieving substantially improved stability.

\begin{figure}
    \centering
    \includegraphics[width=\linewidth]{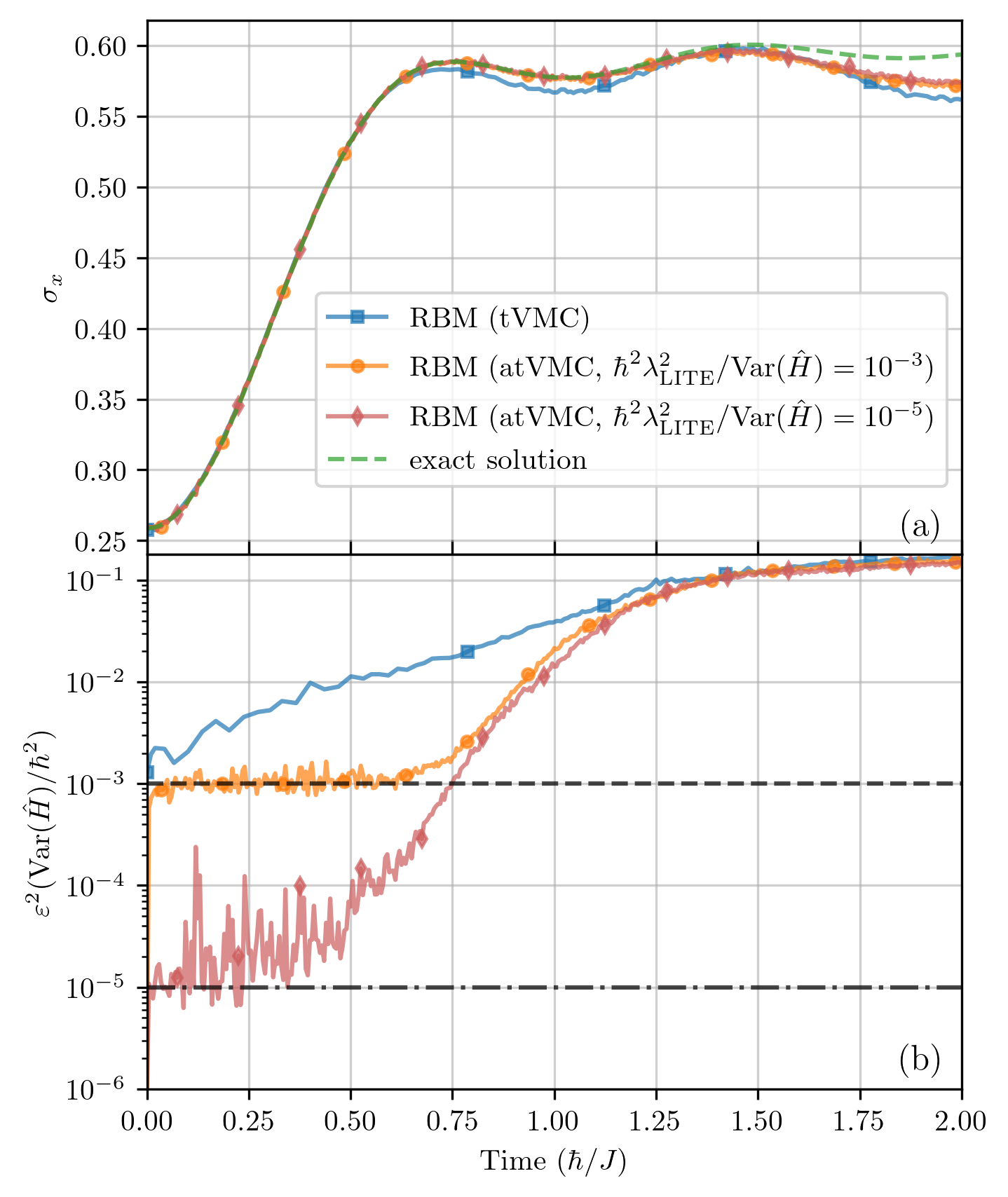}
    \caption{Quench $g=(0.5 \to 1)$ simulated using a RBM ansatz with density $d=15$ on a TFI chain of 32 spins. Three approaches are compared: standard tVMC with SNR regularization and an adaptive time-step scheme (square markers, blue color), atVMC with $\lambda^2_\text{LITE} = 10^{-3} \cdot \operatorname{Var}(\hat{H})/\hbar^2$ and $\eta_{\text{sig}}^2 = 5 \cdot 10^{-3}$ (circular markers, orange color), and atVMC with $\lambda^2_\text{LITE} = 10^{-5} \cdot \operatorname{Var}(\hat{H})/\hbar^2$ and $\eta_{\text{sig}}^2 = 10^{-2}$ (diamond markers, red color). (a) Transverse magnetization per spin: variational results compared to the exact solution (dashed line). Error bars are omitted as they are comparable in magnitude to the observable's oscillations. (b) Squared LITE during the evolution, compared to the target thresholds $\lambda^2_\text{LITE} = 10^{-3} \cdot \operatorname{Var}(\hat{H})/\hbar^2$ (horizontal dashed line) and $\lambda^2_\text{LITE} = 10^{-5} \cdot \operatorname{Var}(\hat{H})/\hbar^2$ (horizontal dash-dotted line).
    }
    \label{fig:RBM_0.5to1_confronto}
\end{figure}

We now compare the performance of atVMC with that of the standard tVMC algorithm using commonly adopted regularization strategies, employing the same number of Monte Carlo samples per time step (here we use about $7 \cdot 10^4$ Monte Carlo samples per time step). Fig.~\ref{fig:RBM_0.5to1_confronto} contrasts the atVMC results with those obtained from a standard tVMC simulation that employs an adaptive time-step scheme combined with the signal-to-noise ratio (SNR) regularization introduced in Ref.~\cite{Schmitt_and_Heyl}. The SNR threshold is set to $4$, a value specifically tuned to prevent instabilities, with energy conservation serving as our primary diagnostic. For the atVMC simulations, we use pseudoinverse regularization with a tolerance of $10^{-7}$. We note in passing that applying this same pseudoinverse regularization in standard tVMC leads to instabilities for this particular case.

Panel (a) of Fig.~\ref{fig:RBM_0.5to1_confronto} shows that the evolution of $\sigma_x$ obtained using atVMC with $\lambda^2_\text{LITE} = 10^{-3} \cdot \operatorname{Var}(\hat{H})/\hbar^2$ and $\eta_{\text{sig}}^2 = 5 \cdot 10^{-3}$ (triangle-marked curve) more closely follows the exact dynamics than the standard tVMC result (circle-marked curve). Panel (b) plots the squared LITE values, revealing that atVMC consistently maintains lower error levels throughout the simulation.
We remark that the accuracy of standard tVMC could be improved by adopting a less expressive ansatz (e.g., an RBM with lower density), which would require milder regularization and thus introduce smaller biases. However, such a choice would limit the representational power of the variational state, particularly in more challenging dynamical regimes. These results highlight that the atVMC framework enables the use of less invasive regularization, even with highly expressive ans\"atze, and can achieve higher accuracy and stability under equivalent computational effort.

In the next example, we illustrate how the atVMC framework can be employed primarily as a regularization tool in this quench scenario. This is achieved by setting $\lambda_{\text{LITE}}^2$ to a very low value, close to the minimum LITE attainable by the ansatz during the simulation. The square-marked curves in Fig.~\ref{fig:RBM_0.5to1_confronto} correspond to $\eta_{\text{sig}}^2 = 10^{-2}$ and $\lambda^2_\text{LITE} = 10^{-5} \cdot \operatorname{Var}(\hat{H})/\hbar^2$. As shown in Panel (b), the full expressivity of the RBM with density $d = 15$ is barely sufficient to reach this error threshold, and the LITE remains above it for most of the simulation. Nevertheless, choosing a threshold that is not strictly unreachable ensures that a collective parameter update is triggered at the beginning of the simulation, allowing the algorithm to identify an initial subset of relevant parameters. Had the threshold been too low to reach even momentarily, no collective freezing would have satisfied the accuracy condition, and all parameters would have remained active throughout. During the rest of the evolution, since the LITE generally stays above the threshold, the number of active parameters is primarily controlled by the overparameterization prevention scheme, effectively enforcing the maximal expressivity compatible with excluding non-relevant parameters, where relevance is defined by the significance threshold $\eta_{\text{sig}}^2$.

While the improvement in the accuracy of the transverse magnetization $\sigma_x$—seen in Fig.~\ref{fig:RBM_0.5to1_confronto}(a)—is modest, Panel (b) shows that the squared LITE achieved with the lower threshold is consistently below that obtained with the higher threshold $\lambda^2_\text{LITE} = 10^{-3} \cdot \operatorname{Var}(\hat{H})/\hbar^2$ throughout the simulation. These results show that atVMC can serve as an effective regularization scheme when the goal is to maintain the highest expressivity of the ansatz without compromising numerical stability.

\subsection{RBM wave function, quench $g=(4 \to 2)$}
\begin{figure}
    \centering
    \includegraphics[width=\linewidth]{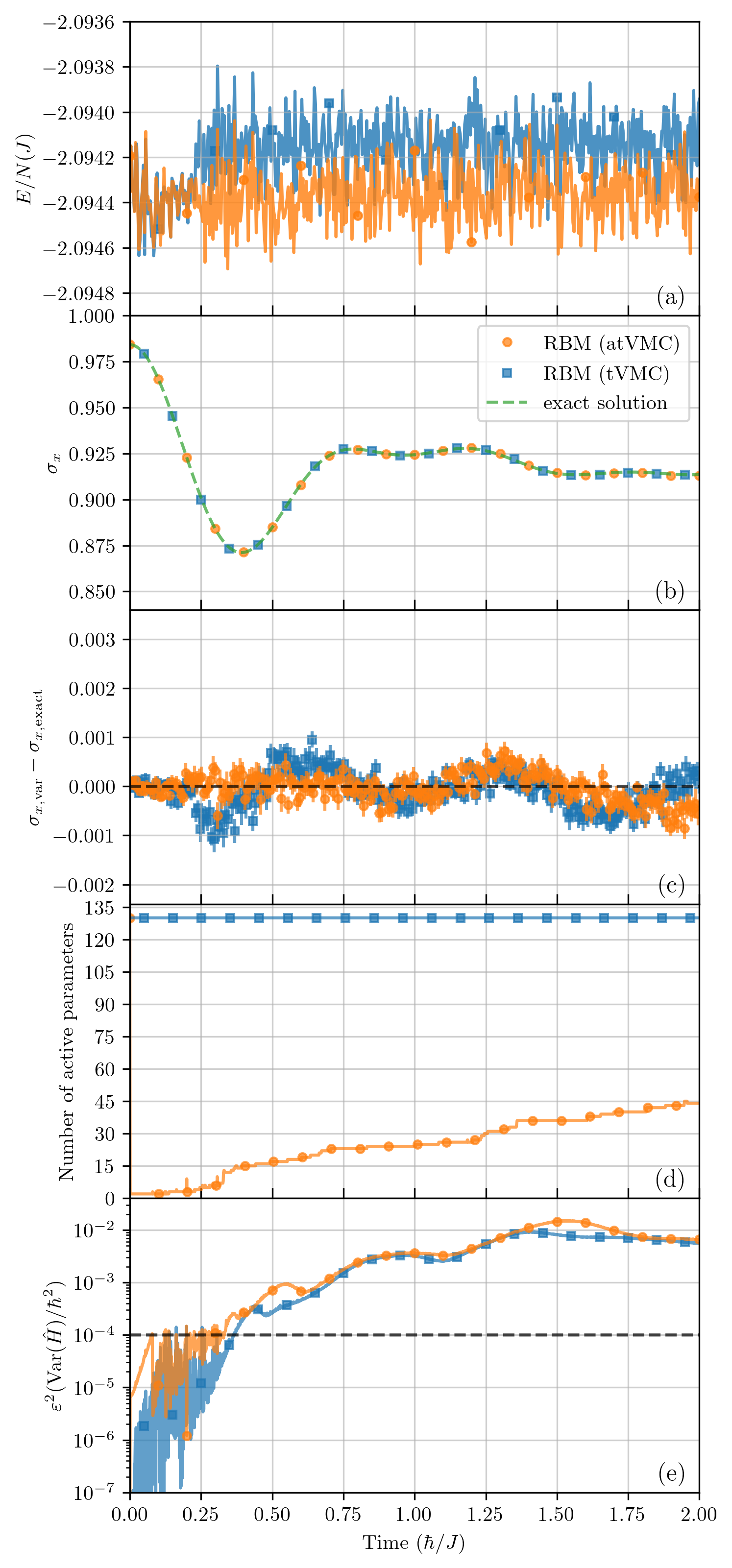}
    \caption{Quench $g=(4 \to 2)$ simulated using atVMC with a RBM ansatz with density $d=1$ on a TFI chain of $128$ spins. Two approaches are compared: standard tVMC (square markers, blue color) and atVMC with $\lambda^2_\text{LITE} = 10^{-4} \cdot \operatorname{Var}(\hat{H})/\hbar^2$ and $\eta_{\text{sig}}^2 = 10^{-2}$ (circular markers, orange color).
    (a) Energy per spin. (b) Transverse magnetization per spin: atVMC and tVMC results compared to the exact solution (dashed line).  Error bars are omitted as they are comparable in magnitude to the observable's oscillations.
    (c) Difference between the variational and the exact transverse magnetization per spin for the atVMC simulation and the tVMC simuation.
    (d) Number of active variational parameters selected by the adaptive algorithm. (e) Squared LITE during the evolution, compared to the target threshold (horizontal dashed line).}
    \label{fig:RBM_4to2_128spin}
\end{figure}
As a final example, we reconsider the quench $g = (4 \to 2)$ in a TFI chain of $128$ spins. 
Increasing the system size also increases the number of variational parameters, so even this relatively simple quench can enter a regime of potential overparameterization. We use a restricted Boltzmann machine (RBM) with density $d=1$, corresponding to 130 variational parameters in our setup.
We set the LITE threshold to $\lambda^2_\text{LITE} = 10^{-4} \cdot \operatorname{Var}(\hat{H})/\hbar^2$  and the significance threshold to $\eta_{\text{sig}}^2 = 10^{-2}$. 
The pseudoinverse regularization is applied with a tolerance of $10^{-7}$.
We compare the atVMC results with a tVMC simulation using the same regularization scheme with identical tolerance.

Fig.~\ref{fig:RBM_4to2_128spin}(a) shows the time evolution of the energy per spin. In the standard tVMC simulation, the applied regularization scheme is not sufficient to suppress minor numerical instabilities, as evidenced by the small jump in Fig.~\ref{fig:RBM_4to2_128spin}(a) around $t\simeq 0.25 \,\hbar/J$. Nevertheless, the atVMC simulation remains stable under the same conditions.
Figures~\ref{fig:RBM_4to2_128spin}(b) and~(c) demonstrate that the variational results are in excellent agreement with the exact evolution, with panel (b) showing the transverse magnetization per spin and panel (c) displaying the corresponding deviations. A small jump in the transverse magnetization obtained from the standard tVMC simulation can be also detected at $t\simeq 0.25 \,\hbar/J$, coinciding with the numerical instability observed for the energy.
In Fig.~\ref{fig:RBM_4to2_128spin}(d) the substantial reduction in the number of evolved variational parameters during the atVMC simulation is evident, while panel (e) shows that the squared LITE remains largely unaffected.

\section{\label{sec:conclusions}Discussion and Conclusions}
We have introduced the adaptive time-dependent variational Monte Carlo (atVMC) algorithm, an extension of the tVMC method, designed to enhance the stability and accuracy of simulations by dynamically adjusting the expressivity of the variational quantum state. The core idea is to estimate each parameter's contribution to the local-in-time error (LITE) using quantities already computed in standard tVMC, enabling the algorithm to evolve only the most relevant parameters while keeping the error below a user-defined threshold. The resulting algorithm supports both individual and collective freezing or unfreezing of parameters, as well as a mechanism to mitigate overparameterization. Together, these features allow atVMC to effectively address the numerical instabilities that commonly arise when using highly expressive variational ans\"atze.

We benchmarked atVMC using quantum quenches in the one-dimensional transverse-field Ising model with variational wave functions of increasing expressivity. In a less challenging quench scenario with the spin-Jastrow ansatz, we showed how atVMC selectively freezes and unfreezes spin-spin correlations in response to entanglement growth, maintaining the LITE below the threshold throughout the evolution. For more challenging quenches using restricted Boltzmann machine (RBM) ans\"atze, we showed the crucial role of collective parameter updates and overparameterization control. These features enabled the algorithm to maintain stability and accuracy in regimes where standard tVMC would require strong regularization. Moreover, the method yields improved agreement with exact results compared to conventional tVMC dynamics constrained by heavy regularization. These advantages make the atVMC strategy a particularly promising enhancement to tVMC for simulations involving highly expressive variational ansätze, where numerical stability is a key challenge.

Although the adaptive strategy can considerably reduce the number of evolving parameters, the overall computational gain is often compensated by the additional effort required to evaluate the relevance of each parameter and determine which should remain active. As a result, the computational cost of atVMC is typically comparable to that of standard tVMC. Even when a substantial fraction of parameters remains frozen for extended portions of the evolution, we did not observe a reduction in wall-clock time exceeding approximately 20\%. The primary advantage of atVMC therefore lies not in raw performance improvement, but in its ability to provide stable and unbiased dynamics in situations where standard tVMC would otherwise require strong regularization or fail to converge.

Looking ahead, several directions could further enhance the scope and effectiveness of atVMC. A natural next step is its application to more complex neural network ans\"atze, where adaptive control could allow the study of complex quantum dynamics with large parameter spaces. Another interesting avenue is the extension of atVMC to the simulation of the dynamics of open quantum systems, where stability and accuracy are equally critical. Finally, allowing the freezing or unfreezing of arbitrary directions in parameter space could offer more refined control over the variational manifold, potentially improving efficiency and accuracy beyond the current parameter-wise scheme.

\section{Acknowledgments}

Numerical simulations were run on computational resources provided by INDACO Platform, which is a project of High Performance Computing at Universit\`a degli Studi di Milano. D.E.G. and C.A. acknowledge support from MUR and Next Generation EU via the PRIN 2022 Project CONTRABASS (Contract N.2022KB2JJM). 

\bibliography{apssamp}

\end{document}